\documentclass[twocolumn,showpacs,a4paper,nofootinbib,tightenlines,floats]{revtex4}
\usepackage{bm}
\usepackage{latexsym}
\usepackage{dcolumn}
\usepackage{amsfonts,amssymb}
\usepackage{graphicx,epsfig}
\usepackage{psfrag}
\usepackage{amsmath,amssymb}
\usepackage{graphicx,epsfig}
\begin{document}
\title{\textbf{Can cosmological observations uniquely determine the nature of dark energy ?}}

\author{Sanil Unnikrishnan}
\email{sanil@physics.du.ac.in}
\email[ ]{sanil.phy@gmail.com}
\affiliation{Department of Physics \& Astrophysics, University of Delhi,
 Delhi : 110007, India}

\date{\today}

\pacs{95.36.+x}

\begin{abstract}
The observational effect of all minimally coupled scalar field models of dark energy can be determined by the behavior of the following  two parameters : (1) equation of state parameter $w$, which relates dark energy pressure to its energy density, and (2) effective speed of sound $c_{e}^{2}$, which relates dark energy pressure fluctuation to its density fluctuation. In this paper we show that these two parameters do not uniquely determine the form of a scalar field dark energy Lagrangian even after taking into account the perturbation in the scalar field. We present this result by showing that two different forms of scalar field Lagrangian can lead to the same values for these paired parameters. It is well known that from the background evolution the Lagrangian of the scalar field dark energy cannot be uniquely determined.  The two models of dark energy presented in this paper are indistinguishable from the evolution of background as well as from the evolution of perturbations from a FRW metric.
\end{abstract}

\maketitle
\section{Introduction}
The known form of matter such as radiation, atoms etc. can only make up $4\%$ of the total matter content of the universe at the present epoch. The nature of the remaining  $96\%$, of which about $23\%$ is dark matter and  $73\%$ is some form of exotic dark energy driving the accelerated expansion of the universe, is  still not completely understood\cite{vs1}. One often invokes scalar fields to fill up the gap of this unknown form of matter especially for dark energy. Scalar field models of dark energy include quintessence\cite{quint}, tachyon\cite{tachyon1,tachyon2}, phantom\cite{phantom,phantom_STG}, k-essence\cite{k-essence} etc. For a detailed review on dark energy see Ref\cite{DEreview}.

The present accelerated expansion of the universe could in fact be an indication of a nonzero value of the cosmological constant\cite{vs2}. 
However, the present cosmological observations neither rule out a cosmological constant nor a scalar field as a candidate for dark energy\cite{vs3}. 
Cosmological constant can be ruled out  if a definitive detection of perturbations in dark energy is made\cite{sanil}.

Dark energy influences cosmological observations such as luminosity distance and angular diameter distance through its effect on the rate of expansion of the universe. For scalar field  models of dark energy, the perturbations in scalar fields affect the evolutions of the  metric perturbations from the FRW metric, which will consequently show up in the ISW effect\cite{DeDeo}. All of these effects of scalar field dark energy on cosmic expansion rate as well as on the ISW effect can be characterized  by two parameters : (1) equation of state parameter given by $w = p/\rho$ and (2) the effective speed of sound $c_{e}^{2}$ which relates pressure fluctuations to density fluctuations. This implies that from cosmological observations one can in principle estimate $w$\cite{vs3,vs4,vs5} and $c_{e}^{2}$\cite{Erickson}. Evolution of the equation of state parameter $w(t)$ alone cannot uniquely determine the Lagrangian of the scalar field dark energy\cite{TRC_paddy}. This implies that from the background evolution $a(t)$ one cannot determine the form of the scalar field Lagrangian uniquely \cite{Feinstein, tachyon1, TRC_paddy, Malquarti}.

The aim of this paper is to investigate whether the form of the scalar field dark energy Lagrangian is uniquely determined if we know the background evolution $a(t)$ as well as the evolution of the  metric perturbation.
In other words the question we are addressing in this paper is  whether the values of $w$ and $c_{e}^{2}$ uniquely fix the form of the scalar field dark energy Lagrangian?
We show that the answer to this is no. We demonstrate this by showing that  two different forms of scalar field Lagrangian given by $(1)$  $ \mathcal{L}_{1} = X^{\alpha} - V_{1}(\phi)$ and  $(2)$  $ \mathcal{L}_{2} =  - V_{2}(\phi)( 1 - 2X)^{\beta}$, where $\alpha$ and $\beta$ are constants and $X = (1/2)\partial_{\mu}\phi\partial^{\mu}\phi$, can lead to the same values of $w$ and $c_{e}^{2}$. This implies that the evolution of the background as well as the metric perturbation  is identical in both of these models. This is achieved by appropriately  choosing the value  of $\beta$ for a given value of $\alpha$. We illustrate with this example that if the present accelerated expansion of the universe is not due to a cosmological constant or  quintessence then it will be impossible to uniquely determine the nature of dark energy from cosmological observations.

In this paper we work in the longitudinal gauge. This paper is organized in the following way : In Sec. \ref{sec :: DE parameters} we show that the evolution of the scale factor $a(t)$ and metric perturbation $\Phi(\vec{x},t)$ in the longitudinal gauge is determined by equation of state parameter $w$ and the effective speed of sound $c_{e}^{2}$ of dark energy. In Sec. \ref{sec :: Generalized Quintessence} we present a model of generalized quintessence dark energy. In section \ref{sec :: Generalized Tachyon} we present a model of generalized tachyon dark energy which influences cosmological observations in exactly the same way as that by the model presented in Sec. \ref{sec :: Generalized Quintessence}. Sec. \ref{sec :: conclusions} summarizes  the results. In addition,  we present  in the appendix a generalized closed set of cosmological perturbation equations applicable to perfect fluid and scalar fields.
In this paper we work in natural units defined as $\hbar = c = 1$.

\section{Dark energy  parameters  $w$ and $c_{e}^{2}$}\label{sec :: DE parameters}
We shall consider a universe with minimally coupled pressureless matter and scalar field dark energy with Lagrangian $\mathcal{L} = \mathcal{L}(X,\phi)$ which is a general function of the kinetic term $X = (1/2)\partial_{\mu}\phi\partial^{\mu}\phi$ and the field $\phi$.
For this system, scalar metric perturbation in the longitudinal gauge is given by \cite{bardeen PRD 1980, Kodama, mukhanov 1992}:
\begin{equation}
    ds^{2} = \left( 1 + 2\Phi\right)dt^{2} - a^{2}(t)\left(1 - 2 \Phi\right)\left[dx^2 + dy^2 +
    dz^2\right]\label{eqn::longitudinal gauge}
\end{equation}
We assume a $k = 0$ (flat) universe.
The evolution of the scale factor $a(t)$ is determined by the following Friedmann equation :
\begin{equation}
\frac{\dot{a}^2}{a^2} = \frac{8 \pi G }{3}\left[\rho_{m0}a^{-3} +  \bar{\rho}_{de}(a)\right]\label{eqn::background eqn 1}
\end{equation}
where $\rho_{m0}$ is the density of the pressureless matter at the present epoch and $\bar{\rho}_{de}(a)$ is the background dark energy density\footnote{All the variable denoted with an over bar such as $\bar{\rho}$ and  $\bar{p}$  corresponds to their average value on the background space time  $ds^{2} = dt^{2} - a^{2}(t)\left[dx^{2} + dy^{2} + dz^{2}\right]$} given by :
\begin{equation}
\bar{\rho}_{de}(a) = \rho_{de0}\exp\left[-3\int ( 1 + w)\frac{da}{a}\right]\label{eqn :: DE density}
\end{equation}
where $\rho_{de0}$ is the homogeneous component of the dark energy density  at the present epoch and $w$ is the equation of state parameter of the dark energy which can be determined by the Lagrangian of the scalar field. This is given by :
\begin{equation}
w = \frac{\bar{\mathcal{L}}(\bar{X},\bar{\phi})}{2\frac{\partial\bar{\mathcal{L}}}{\partial\bar{X}}\bar{X} - \bar{\mathcal{L}}(\bar{X},\bar{\phi})}
\end{equation}
where $\bar{X} = (1/2)\dot{\bar{\phi}}^{2}$ and $\bar{\mathcal{L}} = \bar{\mathcal{L}}(\bar{X},\bar{\phi})$ is the Lagrangian of the background field\footnote{Perturbation in the scalar field is defined as $\phi(\vec{x},t) = \bar{\phi}(t) + \delta\phi(\vec{x},t)$} $\bar{\phi}$ which is only a function  of time. For example we have for a canonical scalar field $\bar{\mathcal{L}}(\bar{X},\bar{\phi}) = (1/2)\dot{\bar{\phi}}(t)^{2} - V(\bar{\phi})$.
The evolution of the scalar field whose dynamics is described by the Lagrangian $\bar{\mathcal{L}}(\bar{X},\bar{\phi})$ is determined by the field equation :
\begin{eqnarray}
\left[\frac{\partial\bar{\mathcal{L}}}{\partial\bar{X}}\dot{\bar{\phi}}\right]^{\textbf{.}} + 3H\left(\frac{\partial\bar{\mathcal{L}}}{\partial\bar{X}}\right)\dot{\bar{\phi}} + \frac{\partial\bar{\mathcal{L}}}{\partial\bar{\phi}} = 0
\end{eqnarray}

Fractional perturbation in the matter density and in dark energy are defined as  :
\begin{eqnarray}
\delta_{m} &\equiv& \frac{\delta\rho_{m}}{\bar{\rho}_m}\\
\delta_{de} &\equiv& \frac{\delta\rho_{de}}{\bar{\rho}_{de}}
\end{eqnarray}
The evolution of metric perturbation $\Phi$, matter perturbation $\delta_{m}$ and perturbation in dark energy $\delta_{de}$  is determined by the following set of  equations :
\begin{eqnarray}
3\frac{\dot{a^2}}{a^2} \Phi + 3 \frac{\dot{a}}{a}\dot{\Phi}+
\frac{k^2\Phi}{a^2} = -4\pi G\left[\rho_{mo}a^{-3} \delta_{m} +  \bar{\rho}_{de}(a) \delta_{de} \right]\label{eqn :: LE1}
\end{eqnarray}
\begin{eqnarray}
\dot{\delta}_{m} &=&  k^{2}u_{m} + 3\dot{\Phi}\\\label{eqn :: PE1}
\dot{u}_{m} &=& -2Hu_{m} - \frac{\Phi}{a^{2}}\\\label{eqn :: PE2}
\dot{\delta}_{de} &=&  \left(1 + w\right)k^{2}u_{de} +  3H\left(w - c_{e}^{2}\right)\delta_{de} + 9H^{2}\nonumber\\
 &&   \times\left( 1 + w\right)\left[c_{e}^{2} - c_{a}^{2}\right]a^{2}u_{de} +  3\left(1 + w\right)\dot{\Phi}\\\label{eqn :: PE3}
 \dot{u}_{de} &=& -H\left(2 - 3c_{e}^{2}\right)u_{de} - \frac{c_{e}^{2}\delta_{de}}{a^{2}\left(1 + w\right)}    -  \frac{\Phi}{a^{2}}\label{eqn :: PE4}
\end{eqnarray}
where $H \equiv \dot{a}/a$,
\begin{eqnarray}
c_{a}^{2} = \frac{\dot{\bar{p}}_{de}}{\dot{\bar{\rho}}_{de}} = w - \frac{\dot{w}}{3H\left(1 + w\right)\bar{\rho}_{de}}
\end{eqnarray}
is the adiabatic sound speed and $c_{e}^{2}$ is the effective sound speed given by \cite{Garriga_mukhanov_1999} :
\begin{equation}
c_{e}^{2} = \frac{\frac{\partial\bar{\mathcal{L}}}{\partial\bar{X}}}{\frac{\partial\bar{\mathcal{L}}}{\partial\bar{X}} + 2 \bar{X}\frac{\partial^{2}\bar{\mathcal{L}}}{\partial\bar{X}^{2}}}\label{eqn : effective sound speed dark energy}
\end{equation}
The effective speed of sound $c_{e}^{2}$ relates dark energy pressure fluctuation to its density fluctuation in the following way \cite{Hu_2004, Gordon_Hu_2004} :
\begin{eqnarray}
\delta p_{de} = c_{e}^{2} \delta \rho_{de} -3H\left(\bar{\rho}_{de} + \bar{p}_{de}\right)a^{2}u_{de}\left[c_{e}^{2} - c_{a}^{2}\right]\label{eqn  :: delta p rho}
\end{eqnarray}

Eq.(\ref{eqn :: LE1}) is  the time-time component of the linearized Einstein equation $\delta G^{\mu}_{\hspace{0.1cm}\nu} = \kappa \delta T^{\mu}_{\hspace{0.1cm}\nu}$. Eqs.(\ref{eqn :: PE1}) to (\ref{eqn :: PE4})  follow from the covariant conservation equation $T^{\mu}_{\hspace{0.2cm}\nu\hspace{0.1cm} ; \hspace{0.1cm}\mu} = 0$,  which are individually valid for both matter and dark energy since they are minimally coupled. In Eqs.(\ref{eqn :: PE1}) to (\ref{eqn :: PE4}), $u_{m}$ and $u_{de}$ are the potential for the respective  peculiar  velocity\footnote{$u_{m}$ and  $u_{de}$ are defined in Appendix}(or velocity perturbation)  $\delta{u}^{i}$ in the perturbed energy momentum tensor $\delta T^{\mu}_{\hspace{0.2cm}\nu} $.

In the gauge in which  $u_{de} = B = 0$, where $B$ is the scalar metric perturbation corresponding to $\delta g_{0i}$, the effective  speed of sound of dark energy is the ratio of its pressure fluctuation to its density fluctuation. For scalar fields, in general, $c_{e}^{2} \neq c_{a}^{2}$. However, for perfect fluids these two sound speeds coincides. The fact  that $c_{e}^{2}$ does not (in general) coincide with $c_{a}^{2}$ is the consequence of non zero
intrinsic entropy perturbation of the scalar field.

For a given value of the equation of state parameter $w$ (which could in general be a function of epoch), Eq.(\ref{eqn::background eqn 1}) and Eq.(\ref{eqn :: DE density}) can be solved to determine the evolution of the scale factor $a(t)$. The perturbation equations [Eqs.(\ref{eqn :: LE1}) to (\ref{eqn :: PE4})] are affected by both $w$ and $c_{e}^{2}$. This implies that $w$ and $c_{e}^{2}$ are the two parameters of the scalar field dark energy which determines the solution $a(t)$ and $\Phi(\vec{x}, t)$ in the line element (\ref{eqn::longitudinal gauge}). The question we are addressing in this paper is whether  two different forms of the Lagrangian $\mathcal{L}(X,\phi)$ lead to the same set of  $w$ and $c_{e}^{2}$. We will argue that this is indeed true by showing that two different forms of the Lagrangian namely generalized quintessence with $ \mathcal{L}_{1} = X^{\alpha} - V_{1}(\phi)$ and generalized tachyon with $ \mathcal{L}_{2} =  - V_{2}(\phi)\left( 1 - 2X\right)^{\beta}$  lead to the same $w$ and $c_{e}^{2}$.
In this paper we only consider the case when both $w$ and $c_{e}^{2}$ are constant.

\section{Generalized Quintessence Dark Energy}\label{sec :: Generalized Quintessence}
We will first consider  a generalized quintessence model of dark energy with Lagrangian given by \cite{Fang}:
\begin{eqnarray}
 \mathcal{L}_{1} = X^{\alpha} - V_{1}(\phi)\label{eqn : lagrangian generalised quientessene}
\end{eqnarray}
where $\alpha$ is a constant. If $\alpha = 1$, then this Lagrangian corresponds to the canonical scalar field or quintessence dark energy \cite{quint}.

We shall reconstruct the form of the potential $V_{1}(\phi)$ such that it leads to the solution for which $w =$ constant. This would then imply that $\bar{\rho}_{de}(a) = \bar{\rho}_{de0}a^{-3(1 + w)}$, where $\bar{\rho}_{de0}$ is the dark energy density at the present epoch. The Friedmann equation [Eq.(\ref{eqn::background eqn 1})] would then becomes :
\begin{eqnarray}
H^{2} = H^{2}_{0}\left[\Omega_{mo}a^{-3} + \Omega_{de0}a^{-3(1 + w)}\right]\label{eqn :: H}
\end{eqnarray}
where $H_{0}$ is the Hubble parameter at the present epoch, $\Omega_{mo} = 8 \pi G \bar{\rho}_{m0}/(3 H^{2}_{0})$ and $\Omega_{deo} = 8 \pi G \bar{\rho}_{de0}/(3 H^{2}_{0})$ are the dimensionless density parameters at the present epoch of matter and dark energy respectively.

From the Lagrangian (\ref{eqn : lagrangian generalised quientessene}), it follows that :
\begin{eqnarray}
\bar{\rho}_{de} &=& \left(\frac{2\alpha - 1}{2^{\alpha}}\right)\dot{\bar{\phi}}^{2\alpha} +  V_{1}(\bar{\phi})\label{eqn :: rho GQ}\\
\bar{p}_{de} &=& \frac{\dot{\bar{\phi}}^{2\alpha}}{2^{\alpha}} -  V_{1}(\bar{\phi})\label{eqn :: p GQ}
\end{eqnarray}

From Eqs.(\ref{eqn :: H}), (\ref{eqn :: rho GQ}) and (\ref{eqn :: p GQ}) we obtain :
\begin{eqnarray}
\frac{d\bar{\phi}}{da} &=& \sqrt{2}\left[\frac{3H_{0}^{2(1 - \alpha)}M_{p}^{2}\left(1 + w\right)\Omega_{deo}}{2\alpha}\right]^{\frac{1}{2\alpha}}\nonumber\\
&& \times\frac{a^{-\lambda}}{\sqrt{\Omega_{mo}a^{-3} + \Omega_{de0}a^{-3(1 + w)}}}\label{eqn :: dphi by da GQ}
\end{eqnarray}
\begin{eqnarray}
V_{1}(a) = 3\left[\frac{1 - (2\alpha - 1)w}{2\alpha}\right]\frac{H_{0}^{2}M_{p}^{2}\Omega_{de0}}{a^{3\left(1 + w\right)}}\label{eqn :: V(a) GQ}
\end{eqnarray}
where $M_{p} = 1/\sqrt{8 \pi G}$ is the Planck mass and
\begin{eqnarray}
\lambda = \frac{3\left(1 + w\right) + 2\alpha}{2\alpha}
\end{eqnarray}

\begin{figure}
\begin{center}
\vskip 0.1in
\resizebox{2.5in}{1.5in}{\includegraphics{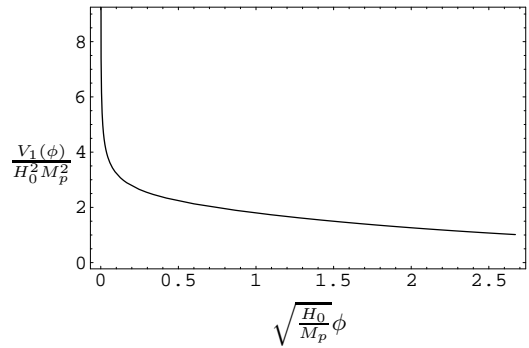}}
\vskip -0.8in  \hskip -2.7in  $\frac{V_{1}(\phi)}{H_{0}^{2}M_{p}^{2}}$
\vskip 0.6in \hskip 0.1in $\sqrt{\frac{H_{0}}{M_{p}}}\phi$
\vskip 0.1in
\caption{The behavior  of the generalized quintessence potential $V_{1}(\phi)$ with $\phi$ for $ w = - 0.9$ and for $\alpha = 2$. For this model $c_{e}^{2} = 1/3$.}\label{fig : GQ}
\end{center}
\end{figure}

In scalar field dominated universe with $\Omega_{de0} = 1$, Eq.(\ref{eqn :: dphi by da GQ})  can be analytically solved to obtain the solution $a(\phi)$.  Substituting this solution in Eq.(\ref{eqn :: V(a) GQ}) leads to the following form of the potential :
\begin{eqnarray}
V_{1}(\phi) = \frac{V_{0}}{\phi^{n}}\label{eqn :: potential GQ}
\end{eqnarray}
where
\begin{eqnarray}
n =\frac{2\alpha}{\alpha - 1}
\end{eqnarray}

This form of the potential [Eq.(\ref{eqn :: potential GQ})] corresponds  to the case when $\alpha \neq 1$. For $\alpha = 1$, which corresponds to standard scalar field or quintessence, Eqs.(\ref{eqn :: dphi by da GQ}) and (\ref{eqn :: V(a) GQ}) lead to an exponential form of the potential in the scalar field dominated universe. In this paper we will only consider the case when the parameter $\alpha \neq 1$.

A realistic model of the universe consistent with observations would require $\Omega_{mo} = 0.27$, $\Omega_{deo} = 0.73$ and $w$ close to minus one \cite{WMAP5, spergel_2006, spergel_2003}. We numerically obtain the form of the potential from Eqs.(\ref{eqn :: dphi by da GQ}) and (\ref{eqn :: V(a) GQ}) with these values for $\Omega_{mo}$ and $\Omega_{deo}$. This is shown in Fig.\ref{fig : GQ} and it corresponds to the same class of the potentials described in Eq.(\ref{eqn :: potential GQ}).
This form of the potential  leads to a solution for which the equation of state parameter $w = - 0.9$  in a universe with pressureless matter and scalar field dark energy with Lagrangian of the form given by  Eq.(\ref{eqn : lagrangian generalised quientessene}) and with the value of the parameter $\alpha = 2$. Eq.(\ref{eqn :: dphi by da GQ}) has been numerically integrated with the initial condition $\sqrt{H_{0}/M_{p}}\phi = 10^{-4}$ at $a = 10^{-3}$.

Since the equation of state parameter $w$ in this model is constant, the adiabatic sound speed $c_{a}^{2} = w$ and the effective sound speed defined in  Eq.(\ref{eqn : effective sound speed dark energy}) is given by \cite{vikman1}:
\begin{eqnarray}
c_{e}^{2} = \frac{1}{2\alpha - 1}\label{eqn :: ce alpha}
\end{eqnarray}

Given any value of $c_{e}^{2}$, assuming it to be constant, we can appropriately choose $\alpha$ given by Eq.(\ref{eqn :: ce alpha}).
For $\alpha > 1$, the effective speed of sound is less than the speed of light \textit{i.e.} $c_{e}^{2} < 1$\cite{Ellis}.
However, k-essence dark energy in general admit solutions for which the speed of sound is greater than the speed of light \cite{Bonvin2, Kang}. The question of whether such superluminal propagation of perturbation on classical background is acceptable or not is  debated in the literature (see for instance Refs.\cite{Bonvin1, Bonvin2, vikman2, Ellis}).

\section{Generalized Tachyon Dark Energy}\label{sec :: Generalized Tachyon}

In this section we will consider a model  of generalized tachyon dark energy with Lagrangian of the form :
\begin{eqnarray}
\mathcal{L}_{2} =  - V_{2}(\phi)\left( 1 - 2X\right)^{\beta}\label{eqn : lagrangian generalised tachyon}
\end{eqnarray}
 For $\beta = 1/2$, this form of the Lagrangian would become the usual DBI form of the Lagrangian \cite{tachyon1, tachyon2}. A model of generalized tachyon field with constant potential  can be found in Refs.\cite{GTF1,GTF2}. In this paper, we will reconstruct  a form of the potential $ V_{2}(\phi)$ such that it leads to the solution for which the equation of state parameter is constant. The value of the parameter $\beta$ would then be fixed such that the effective speed of sound $c_{e}^{2}$ for both  forms of the Lagrangian given by Eq.(\ref{eqn : lagrangian generalised quientessene}) and Eq.(\ref{eqn : lagrangian generalised tachyon}) is exactly the same.

From the Lagrangian given by Eq.(\ref{eqn : lagrangian generalised tachyon}) we obtain :
\begin{eqnarray}
\bar{\rho}_{de} &=& V_{2}(\bar{\phi})\left[\left(2\beta -1\right)\dot{\bar{\phi}}^{2} + 1\right]\left(1 - \dot{\bar{\phi}}^{2}\right)^{\beta -1}\label{eqn :: rho GT}\\
\bar{p}_{de} &=& -V_{2}(\bar{\phi})\left(1 - \dot{\bar{\phi}}^{2}\right)^{\beta}\label{eqn :: p GT}
\end{eqnarray}

For a constant value of the equation of state parameter $w$, Eqs.(\ref{eqn :: rho GT}) and (\ref{eqn :: p GT}) leads to the following equations :

\begin{eqnarray}
\frac{d\bar{\phi}}{da} &=& \frac{1}{H_{o}}\sqrt{\frac{1 + w}{1 - w\left(2\beta - 1\right)}}\nonumber\\
&&\times\frac{1}{a\sqrt{\Omega_{mo}a^{-3} + \Omega_{de0}a^{-3\left(1 + w\right)}}}\label{eqn :: dphi by da GT}
\end{eqnarray}
\begin{eqnarray}
V_{2}(a) = -3w\left[\frac{w\left(2\beta - 1\right) - 1}{2w\beta}\right]^{\beta}\frac{H^{2}_{0}M_{p}^{2}\Omega_{de0}}{a^{3\left(1 + w\right)}}\label{eqn :: V(a) GT}
\end{eqnarray}

\begin{figure}
\begin{center}
\vskip 0.1in
\resizebox{2.5in}{1.5in}{\includegraphics{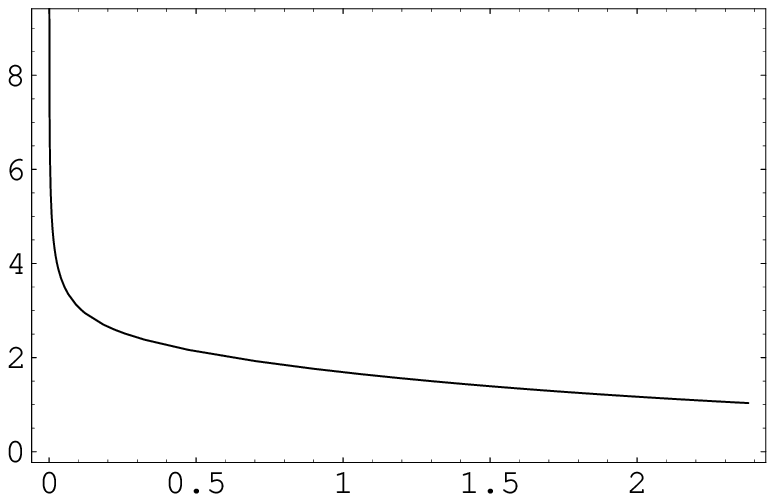}}
\vskip -0.8in  \hskip -2.7in  $\frac{V_{2}(\phi)}{H_{0}^{2}M_{p}^{2}}$
\vskip 0.6in \hskip 0.1in $H_{0}\phi$
\vskip 0.1in
\caption{A form of the generalized tachyon potential $V_{2}(\phi)$ for $ w = - 0.9$ and for $\beta = 1/19$. For this model $c_{e}^{2} = 1/3$.}\label{fig : GT}
\end{center}
\end{figure}

In a scalar field dominated universe with $\Omega_{de} = 1$, Eqs.(\ref{eqn :: dphi by da GT}) and (\ref{eqn :: V(a) GT}) lead to the following form of the potential :
\begin{eqnarray}
V_{2}(\phi) = \frac{V_{0}}{\phi^{2}}\label{eqn :: potential GT}
\end{eqnarray}
This form of the potential with Lagrangian of the form given by Eq.(\ref{eqn : lagrangian generalised tachyon}) leads to a solution in the scalar field dominated universe for which the equation of state parameter is constant. However, for a realistic model of the universe to be consistent with the observation would require that $\Omega_{mo} = 0.27$, $\Omega_{deo} = 0.73$. In this case we numerically obtain the form of the potential from Eqs.(\ref{eqn :: dphi by da GT}) and (\ref{eqn :: V(a) GT}). The form of the potential thus obtained is shown in Fig. \ref{fig : GT}. Eq.(\ref{eqn :: dphi by da GQ}) has been numerically integrated with the initial condition $H_{0}\phi = 10^{-4}$ at $a = 10^{-3}$.

From the Lagrangian Eq.(\ref{eqn : lagrangian generalised tachyon}), the effective speed of sound defined in Eq.(\ref{eqn : effective sound speed dark energy}) would become :
\begin{eqnarray}
c_{e}^{2} = \frac{-\beta w}{(1 - \beta) + w(1 - 2\beta)}
\end{eqnarray}

For a given value of the parameter $\alpha$ in the Lagrangian (\ref{eqn : lagrangian generalised quientessene}) and for a given value of the equation of state parameter $w$, if we choose the value of the parameter $\beta$ such that:
\begin{eqnarray}
\beta = \frac{1 + w}{1 + w(3 - 2\alpha)}\label{eqn :: beta}
\end{eqnarray}
 then the effective speed of sound for both  forms of the Lagrangian is exactly the same. For example if $\alpha = 2$ and $w = -0.9$ then we require  $\beta = 1/19$. For $w = -0.8$ and $\alpha = 2$, the required value of the parameter is $\beta = 1/9$.

 Since the equation of state parameter $w$ for both  models of dark energy presented in this paper is the same, its effect on the cosmic expansion rate $a(t)$ determined by Eqs.(\ref{eqn::background eqn 1}) and (\ref{eqn :: DE density}) would be identical. Also, since the effective speed of sound $c_{e}^{2}$ is the same for both models, its effect on the metric perturbation $\Phi(t, \vec{x})$ as well as on the matter power spectrum determined by Eqs.(\ref{eqn :: LE1}) to (\ref{eqn :: PE4}) would be identical. Hence, if we fix the value of the parameter $\beta$ using Eq.(\ref{eqn :: beta}), then for a given $w$ and $\alpha$, the two forms of the Lagrangian would be observationally indistinguishable.

 If $c_{e}^{2} = 1$, then it turns out from Eqs.(\ref{eqn :: ce alpha}) and  (\ref{eqn :: beta}) that $\alpha = \beta = 1$. Only in this case the two form of the Lagrangian given by  Eqs.(\ref{eqn : lagrangian generalised quientessene}) and (\ref{eqn : lagrangian generalised tachyon}) would be related through a redefinition of the fields. Also the degeneracy in the model space does not hold if the present accelerated expansion is driven by the cosmological constant for which $w = -1$.

  Hence, from cosmological observations we might be able to determine the value of the dark energy parameters $w$ and $c_{e}^{2}$. However, with these values of $w$ and $c_{e}^{2}$ we will not be able to determine uniquely the form of the scalar field Lagrangian if $w \neq -1$ and $c_{e}^{2} \neq 1$. Hence, we emphasize that besides confronting models of dark energy with cosmological observations, we must also devise some mechanism ( or experiments) to determine its nature directly. For example, a model of dark energy described in Ref.\cite{Stojkovic} can likely be tested at the  LHC.

\section{Conclusions}\label{sec :: conclusions}
It is demonstrated  in Refs.\cite{Feinstein, tachyon1, TRC_paddy, Malquarti} that a given background evolution $a(t)$ can be obtained from fundamentally  different forms of the scalar field  Lagrangian.
In this paper we have discussed two scalar field models which are indistinguishable not only from the background evolution $a(t)$ but also from the evolution of metric perturbation $\Phi(t, \vec{x})$ in the longitudinal gauge. We have demonstrated this  by showing that two different models of scalar field dark energy can lead to the same set of two parameters $w$ and $c_{e}^{2}$.

 In this paper we have considered two different models of dark energy with a Lagrangian of the form $ \mathcal{L}_{1} = X^{\alpha} - V_{1}(\phi)$ and  $ \mathcal{L}_{2} =  - V_{2}(\phi)( 1 - 2X)^{\beta}$. We have reconstructed the form of the two potentials $V_{1}(\phi)$ and $ V_{2}(\phi) $ such that in both models $w = - 0.9$. The two constants $\alpha$ and $\beta$ were fixed such that in both  models $c_{e}^{2} = 1/3$. In fact, from  these two forms of the Lagrangian it is possible to reconstruct a  model of dark energy with any value of $w > -1$ and $c_{e}^{2} < 1$, assuming that both $w$ and $c_{e}^{2}$ are constant.

 A universe with roughly about $27\%$ dark matter with negligible pressure and $73\%$ scalar field dark energy with either of the above two Lagrangians will lead to the same solution for the scale factor $a(t)$ and for the metric perturbations about FRW metric, for the same set of initial conditions. Hence, the observable effects of these two models of dark energy will be identical. With this example, we conclude that \emph{it is impossible to uniquely determine the nature of dark energy from cosmological observations if  $w \neq -1$ and $c_{e}^{2} \neq 1$} (for constant $w$ and $c_e^2$).

All of these results emphasize the fact that besides indirectly determining the nature of dark energy through its effect on the cosmic expansion rate, matter power spectrum, ISW effect, etc., we must also devise some mechanism (or experiments) to determine its nature directly.

In this paper we have only considered the case when both $w$ and $c_{e}^{2}$ are constant. The  generalization of this result for $w = w(t)$ and $c_{e}^{2} = c_{e}^{2}(t)$, \textit{i.e} when both $w$ and $c_{e}^{2}$ are function of epoch, is in progress.


\acknowledgements

I am grateful to T. R. Seshadri for useful discussions and for encouraging me to write this paper.
I thank T. Padmanabhan, J. S. Bagla, L. Sriramkumar and H. K. Jassal for discussions.
I also thank C.S.I.R, Govt. of India for senior research fellowship.
\appendix
\section{Cosmological Perturbation equations for perfect fluid/ scalar fields}\label{App :: cosmological perturbation eqns}
In this appendix, we shall present a closed set of cosmological perturbation equations applicable to perfect fluids  and scalar fields.

Scalar metric perturbations describing a perturbed spatially flat FRW line element is given by \cite{bardeen PRD 1980, Kodama, mukhanov 1992} :
\begin{eqnarray}
ds^{2} &=& \left(1 + 2A\right)dt^{2} -2aB_{,i}dx^{i}dt\nonumber\\
 && - a^{2}\left[(1 - 2 \psi)\delta_{ij} + 2E_{,ij}\right]dx^{i}dx^{j}\label{eqn A :: perturbed  FRW line element}
\end{eqnarray}
where $A$ , $\psi$ , $B$ and $E$ are $3-$space scalars.

For most content of the universe such as perfect fluid and  scalar fields the energy momentum tensor can be expressed as :
\begin{equation}
T^{\mu}_{\hspace{0.2cm}\nu} = \left(\rho + p\right)u^{\mu}u_{\nu} - p\delta^{\mu}_{\hspace{0.2cm}\nu} \label{eqn A :: EM tensor}
\end{equation}
where $\rho$ is the energy density, $p$ is the pressure and $u^{\mu}$ is the four velocity field.

We define the perturbations in the energy density $\rho$,  pressure $p$ and the four velocity field $u^{\mu}$ in the following way :
\begin{eqnarray}
\rho(t,\vec{x}) &=& \bar{\rho}(t) + \delta \rho(t,\vec{x}) \label{eqn A :: density perturbation}\\
p(t,\vec{x}) &=&  \bar{p}(t) + \delta p(t,\vec{x}) \label{eqn A :: pressure perturbation}\\
u^{\mu} &=& \bar{u}^{\mu} + \delta u^{\mu} \label{eqn A :: velocity perturbation}
\end{eqnarray}

where $\bar{u}^{\mu} = [1,0,0,0]$ and since $u^{\mu}u_{\mu} = 1$, it follows that $\delta u_{0} = - \delta u^{0} = A$.

The spatial part of the perturbations in the four velocity field $\delta u^{i}$ is the peculiar velocity which can be written as gradient of a scalar :
\begin{eqnarray}
\delta u^{i} = \delta^{ij}u_{\hspace{0.05cm},\hspace{0.05cm}j}
\end{eqnarray}
which implies that $\delta T^{0}_{\hspace{0.2cm}i} = \left(\bar{\rho} +  \bar{p}\right)u_{,i}$ and
\begin{eqnarray}
\delta T^{i}_{\hspace{0.2cm}0} = - a^{2}\left(\bar{\rho} +  \bar{p}\right)\left[u_{,i} + a^{-1}B_{,i}\right]
\end{eqnarray}

Under infinitesimal coordinate transformation defined as $x^{\mu} \rightarrow \tilde{x}^{\mu} =  x^{\mu} + \xi^{\mu}$, where $\xi^{\mu} = (\xi^{0},  \delta^{ij}\xi_{,j})$,   the variables describing metric perturbations $A$, $\psi$, $B$ and $E$ transforms as :
\begin{eqnarray}
\widetilde{A} &=& A  - \dot{\xi}^{o}\label{eqn A :: gauge transformation phi}\\
\widetilde{\psi} &=& \psi  - H\xi^{o}\label{eqn A :: gauge transformation psi}\\
\widetilde{B} &=& B + a^{-1}\xi^{o} - a\dot{\xi}\label{eqn A :: gauge transformation B}\\
\widetilde{E} &=& E - \xi\label{eqn A :: gauge transformation E}
\end{eqnarray}
Similarly the variables describing matter perturbations $\delta \rho$, $\delta p$ and $u$ transforms as :
\begin{eqnarray}
\widetilde{\delta \rho} &=& \delta \rho - \dot{\rho}_{o}\xi^{o}\label{eqn A :: gauge transformation delta rho}\\
\widetilde{\delta p} &=& \delta p - \dot{p}_{o}\xi^{o}\label{eqn A :: gauge transformation delta p}\\
\widetilde{u} &=& u + \dot{\xi}\label{eqn A :: gauge transformation u}
\end{eqnarray}

Linearized Einstein's Equation $\delta G^{\mu}_{\hspace{0.2cm}\nu} = \kappa \delta T^{\mu}_{\hspace{0.2cm}\nu}$, which relates variables describing metric perturbations ($A$, $\psi$, $B$ and $E$) to the variables describing matter perturbations ($\rho$, $\delta p$ and $u$)  is given by :

\begin{eqnarray}
3H^{2}A + H\dot{\psi} + \frac{k^{2}}{a^{2}}\left[\psi - H\left(aB - a^{2}\dot{E}\right)\right] &=& - 4 \pi G \delta \rho\nonumber\\
 && \label{eqn A :: linearise einstein eqn 1}\\
\ddot{\psi} + 3H\dot{\psi} + H\dot{A} + \left(2\dot{H} + 3H^{2}\right)A &=& 4 \pi G \delta p\nonumber\\
 && \label{eqn A :: linearise einstein eqn 2}
\end{eqnarray}
\begin{eqnarray}
\dot{\psi} + HA  = - 4 \pi G a^{2}\left(\bar{\rho} + \bar{p}\right)\left[u + a^{-1}B\right]\label{eqn A :: linearise einstein eqn 3}
\end{eqnarray}

 In Eq.(\ref{eqn A :: linearise einstein eqn 2})  we have used the fact  that for perfect fluids and for scalar fields anisotropic stress is zero \textit{i.e} $\delta T^{i}_{\hspace{0.2cm}j} \propto \delta^{i}_{j}$ for $i$, $j$ $=$ $1$, $2$, $3$. This implies that :

\begin{eqnarray}
A - \psi + H\left(aB - a^{2}\dot{E}\right) + \left(aB - a^{2}\dot{E}\right)^{\textbf{.}} =  0\label{eqn A :: no anisotropic stress}
\end{eqnarray}

The covariant conservation equation $T^{\mu}_{\hspace{0.2cm}\nu\hspace{0.1cm} ; \hspace{0.1cm}\mu} = 0$ leads to the following equations :
\begin{eqnarray}
\dot{\delta \rho} &=& \left(\bar{\rho} + \bar{p}\right)k^{2}u - 3H\left(\delta \rho + \delta p\right) + \left(\bar{\rho} + \bar{p}\right)\left[3\dot{\psi} + k^{2}\dot{E}\right]\nonumber\\
&&\label{eqn A :: rho dot}\\
\dot{u} &=& -H\left(2 - 3c_{a}^{2}\right)\left[u + \frac{B}{a}\right] - \frac{\delta p}{a^{2}\left(\bar{\rho} + \bar{p}\right)} - \frac{(aB\dot{)}}{a^{2}} - \frac{A}{a^{2}}\nonumber\\
&&\label{eqn A :: u dot1}
\end{eqnarray}

For solving the perturbation equations it is required to know how the fluctuation in pressure $\delta p$ is related to the fluctuation in the energy density $\delta \rho$. This is in general determined by the Lagrangian of the matter field.

If the perturbations are such that uniform density gauge coincide with the uniform pressure gauge, then such perturbations are known as adiabatic perturbations. This means that for adiabatic perturbations, we can choose $\xi^{o}$  in Eqs.(\ref{eqn A :: gauge transformation delta rho}) and (\ref{eqn A :: gauge transformation delta p}) such that in the new gauge $\widetilde{\delta \rho} = \widetilde{\delta p} = 0$. This implies that for  adiabatic perturbations in any arbitrary gauge
\begin{equation}
\delta p = c_{a}^{2}\delta \rho
\end{equation}
where $c_{a}^{2}$ is the adiabatic sound speed given by:
\begin{equation}
c_{a}^{2} = \frac{\dot{\bar{p}}}{\dot{\bar{\rho}}}\label{eqn A :: adiabatic sound speed}
\end{equation}
This relation is true when the universe is dominated by a single perfect fluid but does not in general holds for scalar fields.

In general, the  pressure fluctuation can be described as
\begin{equation}
\delta p = c_{a}^{2}\delta \rho + \delta p_{nad}
\end{equation}
where $\delta p_{nad}$ is the non adiabatic pressure fluctuation which is gauge invariant according to gauge transformations (\ref{eqn A :: gauge transformation delta rho}) and (\ref{eqn A :: gauge transformation delta p}).

\subsection{Effective speed of sound}
In non relativistic fluid mechanics, the speed of sound is given by $c_{s}^{2} = \delta p/\delta \rho$. The sound speed is thus determined by the property of the fluid $p = f(\rho)$.

However, in cosmological perturbation theory, the gauge transformations given by Eqs.(\ref{eqn A :: gauge transformation delta rho}) and (\ref{eqn A :: gauge transformation delta p}) imply that the ratio $\delta p/\delta \rho$ would be gauge dependent. The speed of sound must be gauge invariant so that it is solely determined by the property of the fluid not dependent on the choice of the gauge. We expect that  the speed of sound be the ratio of some form of the gauge invariant pressure fluctuation to the gauge invariant density fluctuation.

It is not possible to  construct gauge invariant $\delta \rho$ and $\delta p$ solely from variables describing matter perturbations $\delta \rho$, $\delta p$ and $u$ using Eqs.(\ref{eqn A :: gauge transformation delta rho}) to (\ref{eqn A :: gauge transformation u}).
 However, if we include gauge transformation for metric perturbation $B$ using Eq.(\ref{eqn A :: gauge transformation B}), then we can construct the following gauge invariant density fluctuation $\delta \rho$ and pressure fluctuation $\delta p$ :
\begin{eqnarray}
^{(gi)}\delta \rho &=& \delta \rho + a \dot{\bar{\rho}}\left[au + B\right]\\
^{(gi)}\delta p &=& \delta p + a \dot{\bar{p}}\left[au + B\right]
\end{eqnarray}
Using this we define the gauge invariant effective sound speed $c_{e}^{2}$ as :
\begin{equation}
c_{e}^{2} = \frac{^{(gi)}\delta p}{^{(gi)}\delta \rho} = \frac{\delta p + a \dot{\bar{p}}\left[au + B\right]}{\delta \rho + a \dot{\bar{\rho}}\left[au + B\right]}\label{eqn A :: effective sound speed}
\end{equation}

Gauge transformations (\ref{eqn A :: gauge transformation B}) and (\ref{eqn A :: gauge transformation u}) allows us to define a gauge in which $ B = u = 0$. This gauge is known as the rest frame gauge because in this gauge the peculiar velocity $\delta u^{i} = 0$ and the perturbed energy momentum tensor  becomes diagonal \textit{i.e} $\delta T^{\mu}_{\hspace{0.2cm}\nu\hspace{0.1cm}} = 0$ for all $\mu \neq \nu $. With the density and the pressure fluctuation in the rest gauge given by $\delta \rho_{rest}$ and $\delta p_{rest}$ respectively, Eq.(\ref{eqn A :: effective sound speed}) implies that $c_{e}^{2} = \delta p_{rest}/\delta \rho_{rest}$. Hence the gauge invariant effective sound speed $c_{e}^{2}$ can be physically interpreted as the sound speed in the rest frame gauge in which the peculiar velocity $\delta u^{i}$ is zero.

For a perfect fluid with adiabatic perturbations, which means that $\delta p = c_{a}^{2}\delta \rho$, Eq.(\ref{eqn A :: effective sound speed}) implies that $c_{e}^{2} = c_{a}^{2}$. Hence for a perfect fluid, there is only one sound speed and that is $c_{a}^{2}$ given by Eq.(\ref{eqn A :: adiabatic sound speed}). However,  this is not true for scalar fields.

\subsection{A general non canonical scalar field}

A general non canonical scalar field $\phi$ has a Lagrangian of the form :
\begin{eqnarray}
\mathcal{L} = \mathcal{L}(X,\phi)\label{eqn A :: lagrangian non canonical scalar field}
\end{eqnarray}
where
\begin{eqnarray}
X = \frac{1}{2}\partial_{\mu}\phi\partial^{\mu}\phi
\end{eqnarray}
For this Lagrangian, the energy momentum tensor is given by :
\begin{eqnarray}
T^{\mu}_{\hspace{0.2cm}\nu} = \frac{\partial\mathcal{L}}{\partial X}\partial^{\mu}\phi\partial_{\nu}\phi - \mathcal{L}\delta^{\mu}_{\hspace{0.2cm}\nu}
\end{eqnarray}
Perturbation in the scalar field is defined as
\begin{eqnarray}
\phi(\vec{x},t) = \bar{\phi}(t) + \delta\phi(\vec{x},t)\label{eqn A :: scalar field perturbation}
\end{eqnarray}
For the background space time we can associate the following density $\bar{\rho}$ and pressure $\bar{p}$ :
\begin{eqnarray}
\bar{\rho} &=& 2\frac{\partial\bar{\mathcal{L}}}{\partial\bar{X}}\bar{X} - \bar{\mathcal{L}}(\bar{X},\bar{\phi})\label{eqn A :: density non canonical scalar field}\\
\bar{p} &=& \bar{\mathcal{L}}(\bar{X}, \bar{\phi})\label{eqn A :: pressure non canonical scalar field}
\end{eqnarray}
where $\bar{X} = \frac{1}{2}\dot{\bar{\phi}}^{2}$ and $\bar{\mathcal{L}} = \bar{\mathcal{L}}(\bar{X}, \bar{\phi})$ is the Lagrangian of the background field obtained by treating the scalar field as a function only of time  in Eq.(\ref{eqn A :: lagrangian non canonical scalar field}). For example, for canonical scalar field $\bar{\mathcal{L}}(\bar{X}, \bar{\phi}) = \frac{1}{2}\dot{\bar{\phi}}^{2} - V(\bar{\phi})$.

Considering the perturbations in the scalar fields defined in Eq.(\ref{eqn A :: scalar field perturbation}), we can associate the following $\delta \rho$, $\delta p$ and $u$ for the scalar field with the Lagrangian of the form Eq.(\ref{eqn A :: lagrangian non canonical scalar field}) :
\begin{eqnarray}
\delta\rho &=& \left(\dot{\bar{\phi}}\dot{\delta\phi} - A\dot{\bar{\phi}}^2\right)\left[\frac{\partial\bar{\mathcal{L}}}{\partial\bar{X}} + 2 \bar{X}\frac{\partial^{2}\bar{\mathcal{L}}}{\partial\bar{X}^{2}}\right]\nonumber \\
 && -  \left[\frac{\partial\bar{\mathcal{L}}}{\partial\bar{\phi}}- 2\bar{X}\frac{\partial^{2}\bar{\mathcal{L}}}{\partial\bar{X}\partial\bar{\phi}}\right]\delta\phi \label{eqn A :: perturbed density NC scalar field}\\
\delta p &=& \left(\dot{\bar{\phi}}\dot{\delta\phi} - A\dot{\bar{\phi}}^2\right)\frac{\partial\bar{\mathcal{L}}}{\partial\bar{X}} +   \frac{\partial\bar{\mathcal{L}}}{\partial\bar{\phi}}\delta\phi \label{eqn A :: perturbed pressure NC scalar field}\\
u &=& -\frac{\delta \phi}{a^{2}\dot{\bar{\phi}}} - \frac{B}{a}\label{eqn A :: peculiar velocity NC scalar field}
\end{eqnarray}

The above equation [Eq.(\ref{eqn A :: peculiar velocity NC scalar field})] implies that $u$ for a Lagrangian of the form Eq.(\ref{eqn A :: lagrangian non canonical scalar field}) is the same as that for a canonical scalar field. Substituting Eqs.(\ref{eqn A :: perturbed density NC scalar field}) to (\ref{eqn A :: peculiar velocity NC scalar field}) in Eq.(\ref{eqn A :: effective sound speed}), we find that for a Lagrangian of the form of Eq.(\ref{eqn A :: lagrangian non canonical scalar field}), effective speed of sound is given by \cite{Garriga_mukhanov_1999} :
\begin{equation}
c_{e}^{2} = \frac{\frac{\partial\bar{\mathcal{L}}}{\partial\bar{X}}}{\frac{\partial\bar{\mathcal{L}}}{\partial\bar{X}} + 2 \bar{X}\frac{\partial^{2}\bar{\mathcal{L}}}{\partial\bar{X}^{2}}}\label{eqn A :: effective sound speed NC scalar field}
\end{equation}

Equation (\ref{eqn A :: effective sound speed NC scalar field}) implies that for canonical scalar fields $c_{e}^{2} = 1$\cite{Hu_1998}.

\subsection{Equations of Perturbation}

One of the gauge in which cosmological perturbations can be studied is the longitudinal gauge defined by $B = E = 0$. In this gauge, for both scalar field and for perfect fluid, Eq.(\ref{eqn A :: no anisotropic stress}) implies that :
\begin{eqnarray}
A_{l} = \psi_{l} \equiv \Phi
\end{eqnarray}
where we have denoted the metric perturbation in the longitudinal gauge by $\Phi$. Using Eq.(\ref{eqn A :: effective sound speed}), we find that in this gauge, the pressure fluctuation $\delta p$ is related to the density fluctuation $\delta \rho$ as :
\begin{eqnarray}
\delta p = c_{e}^{2} \delta \rho -3H\left(\bar{\rho} + \bar{p}\right)a^{2}u\left[c_{e}^{2} - c_{a}^{2}\right]\label{eqn A :: delta p rho}
\end{eqnarray}

This is a general relation between the pressure fluctuation and the density fluctuation. For perfect fluid with constant equation of state parameter $c_{e}^{2} = c_{a}^{2} = w$, which from Eq.(\ref{eqn A :: delta p rho}) implies that $\delta p = w \delta \rho$. However, for scalar fields, in general $c_{e}^{2} \neq c_{a}^{2}$.

The relation Eq.(\ref{eqn A :: delta p rho}) closes the equation of perturbations Eqs.(\ref{eqn A :: linearise einstein eqn 1}), (\ref{eqn A :: rho dot}) and (\ref{eqn A :: u dot1}). Defining the fractional density perturbations as $\delta \equiv \delta \rho/\bar{\rho}$, Eqs.(\ref{eqn A :: rho dot}) and (\ref{eqn A :: u dot1}), together with Eq.(\ref{eqn A :: linearise einstein eqn 1}) in longitudinal gauge would become :
\begin{eqnarray}
\dot{\Phi} &=& - H\Phi - \frac{k^2\Phi}{3Ha^{2}} - \frac{4 \pi G}{3H}\bar{\rho} \delta\label{eqn A :: PE1}\\
\dot{\delta} &=&  \left(1 + w\right)k^{2}u +  3H\left(w - c_{e}^{2}\right)\delta  +   9H^{2} \nonumber\\
 &&\times \left( 1 + w\right)\left[c_{e}^{2} - c_{a}^{2}\right]a^{2}u +  3\left(1 + w\right)\dot{\Phi}\\\label{eqn A :: PE2}
 \dot{u} &=& -H\left(2 - 3c_{e}^{2}\right)u - \frac{c_{e}^{2}\delta}{a^{2}\left(1 + w\right)}    -  \frac{\Phi}{a^{2}}\label{eqn A :: PE3}
\end{eqnarray}

These three equations form a close set of equations if the universe is dominated by a single perfect field or a single scalar field. In case of perfect fluid $c_{e}^{2} = c_{a}^{2}$. However, for scalar fields $c_{e}^{2}$ is given by Eq.(\ref{eqn A :: effective sound speed NC scalar field}).


\begin{thebibliography}{99}

\bibitem{DEreview}E.~J. Copeland, M. Sami, S. Tsujikawa, Int. J. Mod. Phys. D
\textbf{15}, 1753 (2006)[hep-th/0603057].

\bibitem{vs1}V. Sahni, Lect. Notes Phys. \textbf{653}, 141(2004) [arXiv:astro-ph/0403324].

\bibitem{quint}
I.~Zlatev, L.~Wang and P.~J.~Steinhardt, Phys.\ Rev.\ Lett.\ \textbf{82}, 896 (1999) [astro-ph/9807002];
A.~D.~Macorra and G.~Piccinelli,  Phys.\ Rev.\ D\ \textbf{61}, 123503 (2000) [hep-ph/9909459];
L.~A.~Ure{\~ n}a-L{\' o}pez and T.~Matos, Phys.\ Rev.\ D\ \textbf{62}, 081302 (2000) [astro-ph/0003364];
P.~F.~Gonz{\'a}lez-D{\'{\i}}az, Phys.\ Rev.\ D\ \textbf{62}, 023513 (2000) [astro-ph/0004125];
C.~Rubano and P.~Scudellaro, Gen.\ Rel.\ Grav.\ \textbf{34}, 307 (2002) [astro-ph/0103335];
S.~A.~Bludman and M.~Roos, Phys.\ Rev.\ D\ \textbf{65}, 043503 (2002) [astro-ph/0109551];
A.~Albrecht and C.~Skordis, Phys.\ Rev.\ Lett.\ \textbf{84}, 2076 (2000) [astro-ph/9908085].
A.~R.~Liddle and  R.~J.~Scherrer, Phys.\ Rev.\ D\ \textbf{59}, 023509 (1998)[astro-ph/9809272].
Z.~K.~Guo, N.~Ohta and Y.~Z.~Zhang,  Phys.\ Rev.\  D {\bf 72}, 023504 (2005) [astro-ph/0505253].
D.~F.~Mota and C.~van de Bruck, Astron.\ Astrophys.\ \textbf{421}, 71 (2004) [arXiv:astro-ph/0401504].
\bibitem{Feinstein}
A.~Feinstein, Phys.\ Rev.\  D\ \textbf{66}, 063511 (2002) arXiv:hep-th/0204140v2.
\bibitem{tachyon1}
T.~Padmanabhan, Phys.\ Rev.\ D\ \textbf{66}, 021301 (2002) [hep-th/0204150]
 \bibitem{TRC_paddy}
 T. Padmanabhan, T. Roy Choudhury, Mon.\ Not.\ Roy.\ Astron.\ Soc.\ \textbf{344}, 823 (2003) [arXiv:astro-ph/0212573v2]
\bibitem{Malquarti} M.~Malquarti, E.~J.~Copeland, A.~ R.~Liddle, M.~Trodden,
Phys.\ Rev.\ D \textbf{67}, 123503 (2003) arXiv:astro-ph/0302279.

\bibitem{tachyon2}
J.~S.~Bagla, H.~K.~Jassal and T.~Padmanabhan, Phys.\ Rev.\ D\ \textbf{67}, 063504 (2003) [astro-ph/0212198];
H.~K.~Jassal, Pramana\ \textbf{62}, 757 (2004) [astro-ph/0303406];
J.~M.~Aguirregabiria and R.~Lazkoz, Phys.\ Rev.\ D\ \textbf{69}, 123502 (2004) [hep-th/0402190];
A.~Sen, Phys.\ Scripta\ T\ \textbf{117}, 70 (2005) [hep-th/0312153];
V.~Gorini, A.~Kamenshchik, U.~Moschella and V.~Pasquier, Phys.\ Rev.\ D\ \textbf{69}, 123512 (2004) [hep-th/0311111];
G.~W.~Gibbons, Class.\ Quan.\ Grav.\ \textbf{20}, S321 (2003) [hep-th/0301117];
C.~Kim, H.~B.~Kim and Y.~Kim, Phys.\ Lett.\  B\ \textbf{552}, 111 (2003) [hep-th/0210101];
G.~Shiu and I.~Wasserman, Phys.\ Lett.\  B\ \textbf{541}, 6 (2002) [hep-th/0205003];
D.~Choudhury, D.~Ghoshal, D.~P.~Jatkar and S.~Panda, Phys.\ Lett.\  B\ \textbf{544}, 231 (2002) [hep-th/0204204];
A.~Frolov, L.~Kofman and A.~Starobinsky, Phys.\ Lett.\  B\ \textbf{545}, 8 (2002) [hep-th/0204187];
G.~W.~Gibbons, Phys.\ Lett.\  B\ \textbf{537} 1, (2002) [hep-th/0204008];
A.~Das, S.~Gupta, T.~Deep~Saini and S.~Kar, Phys.\ Rev.\ D\ \textbf{72}, 043528 (2005) [astro-ph/0505509];
I.~Y.~Aref'eva, astro-ph/0410443;
G. Calcagni, A. R. Liddle, Phys.\ Rev.\ D\ \textbf{74}, 043528 (2006) [astro-ph/0606003];
 E. J. Copeland, M. R. Garousi, M. Sami and S. Tsujikawa,  Phys.\ Rev.\ D\ \textbf{71}, 043003
  (2005) [hep-th/0411192].
\bibitem{phantom}
R.~R.~Caldwell, Phys.\ Letts.\  B\ \textbf{545}, 23 (2002) [astro-ph/9908168];
J.~Hao and X.~Li, Phys.\ Rev.\ D\ \textbf{68}, 043501 (2003) [hep-th/0305207];
G.~W.~Gibbons, hep-th/0302199;
V.~K.~Onemli and R.~P.~Woodard, Phys.\ Rev.\ D\ \textbf{70}, 107301 (2004) [gr-qc/0406098];
S.~Nojiri and S.~D.~Odintsov, Phys.\ Letts.\  B\ \textbf{562}, 147 (2003) [hep-th/0303117];
S.~M.~Carroll, M.~Hoffman and M.~Trodden,  Phys.\ Rev.\ D\ \textbf{68}, 023509 (2003) [astro-ph/0301273];
P.~Singh, M.~Sami and N.~Dadhich, Phys.\ Rev.\ D\ \textbf{68}, 023522 (2003) [hep-th/0305110];
P.~H.~Frampton, Mod.\ Phys.\ Letts.\ A\ \textbf{19}, 801 (2004) [hep-th/0302007];
J.~Hao and X.~Li, Phys.\ Rev.\ D\ \textbf{67}, 107303 (2003) [gr-qc/0302100];
P.~Gonz{\' a}lez-D{\'{\i}}az, Phys.\ Rev.\ D\ \textbf{68}, 021303 (2003) [astro-ph/0305559];
M.~P.~Dabrowski, T.~Stachowiak and M.~Szyd{\l}owski, Phys.\ Rev.\ D\ \textbf{68}, 103519 (2003) [hep-th/0307128];
J.~M.~Cline, S.~Jeon and G.~D.~Moore, Phys.\ Rev.\ D\ \textbf{70}, 043543 (2004) [hep-ph/0311312];
W.~Fang, H.~Q.~Lu, Z.~G.~Huang, and K.~F.~Zhang, Int.\ J.\ Mod.\ Phys.\ D\ \textbf{15}, 199 (2006) [hep-th/0409080];
S.~Nojiri and S.~D.~Odinstov, Phys.\ Rev.\ D\ \textbf{72},  023003 (2005) [hep-th/0505215];
S.~Nesseris and L.~Perivolaropoulos, Phys.\ Rev.\ D\ \textbf{70}, 123529 (2004) [astro-ph/0410309];
S.~Nojiri and S.~D.~Odinstov, Phys.\ Rev.\ D\ \textbf{70}, 103522 (2004) [hep-th/0408170];
E.~Elizalde, S.~Nojiri and S.~D.~Odinstov, Phys.\ Rev.\ D\ \textbf{70}, 043539 (2004) [hep-th/0405034];
S.~Nojiri, S.~D.~Odinstov and S.~Tsujikawa,  Phys.\ Rev.\ D\ \textbf{71}, 063004 (2005) [hep-th/0501025].
Z.~K.~Guo, N.~Ohta and Y.~Z.~Zhang, Mod.\ Phys.\ Lett.\  A {\bf 22}, 883 (2007) [astro-ph/0603109]
\bibitem{phantom_STG}
B.~Boisseau, G.~Esposito-Farese, D.~Polarski, A.~A.~Starobinsky,
Phys. Rev. Lett. \textbf{85}, 2236 (2000) [gr-qc/0001066]
\bibitem{k-essence}
C.~Armendariz-Picon, V.~Mukhanov and P.~J.~Steinhardt, Phys.\ Rev.\ D\ \textbf{63}, 103510 (2001) [astro-ph/0006373];
T.~Chiba, Phys.\ Rev.\ D\ \textbf{66}, 063514 (2002) [astro-ph/0206298];
M.~Malquarti, E.~J.~Copeland, A.~R.~Liddle and M.~Trodden, Phys.\ Rev.\ D\ \textbf{67}, 123503 (2003) [astro-ph/0302279];
L.~P.~Chimento and A.~Feinstein, Mod.\ Phys.\ Letts.\ A\ \textbf{19}, 761 (2004) [astro-ph/0305007];
R.~J.~Scherrer, Phys.\ Rev.\  Letts.\  \textbf{93}, 011301 (2004) [astro-ph/0402316].
A.~A.~Sen, JCAP \textbf{0603}, 010 (2006) [astro-ph/0512406].
\bibitem{vs2}
V.~Sahni and A.~Starobinsky, Int.\ J.\  Mod.\ Phys.\ D \textbf{9}, 373 (2000) [arXiv:astro-ph/9904398];
V.~Sahni, Class.\ Quant.\ Grav.\ \textbf{19}, 3435 (2002) [arXiv:astro-ph/0202076].
\bibitem{vs3} U.~Alam, V.~Sahni and A.~A.~Starobinsky, JCAP \textbf{0702}, 011 (2007) [arXiv:astro-ph/0612381].
\bibitem{vs4}V.~Sahni and A.~Starobinsky, Int.\ J.\ Mod.\ Phys.\ D \textbf{15},  2105 (2006) [arXiv:astro-ph/0610026]
\bibitem{vs5}T.~D.~Saini, S.~Raychaudhury, V.~Sahni and A.~A.~Starobinsky, Phys.\ Rev.\ Lett.\ \textbf{85}, 1162 (2000) [arXiv:astro-ph/9910231]
\bibitem{sanil}
S.~Unnikrishnan, H.~K.~Jassal and T.~R.~Seshadri, arXiv:0801.2017[astro-ph]
\bibitem{DeDeo}
 S.~DeDeo, R.~R.~Caldwell, P.~J.~Steinhardt,  Phys.\ Rev.\ D \textbf{67}, 103509 (2003) arXiv:astro-ph/0301284.
\bibitem{Erickson}
 J.~K.~Erickson, R.~R.~Caldwell, P.~ J.~Steinhardt, C.~Armendariz-Picon, V.~Mukhanov,
 Phys.\ Rev.\ Lett.\  \textbf{88}, 121301 (2002) arXiv:astro-ph/0112438.
\bibitem{bardeen PRD 1980}
J. Bardeen,  Phys. Rev. D  \textbf{22}, 1882 (1980)
\bibitem{Kodama}
H. Kodama and M. Sasaki, Prog. Theor. Phys. Suppl., \textbf{78}, 1 (1984).
\bibitem{mukhanov 1992}
V.~F.~Mukhanov, H.~A.~Feldman and R.~H.~Brandenberger,  Phys.\ Rep.\ \textbf{215}, 203 (1992).
\bibitem{Garriga_mukhanov_1999}
J.~Garriga and V.~Mukhanov, Phys.\ Lett.\ B\ \textbf{458}, 219 (1999). [hep-th/9904176]
\bibitem{Hu_2004}W.~Hu, Phys.\ Rev.\ D \textbf{71}, 047301 (2005).[astro-ph/0410680]
\bibitem{Gordon_Hu_2004}
C.~Gordon and W.~Hu Phys.\ Rev.\ D \textbf{70} (2004) 083003. [astro-ph/0406496]
\bibitem{Fang}
 W.~Fang, H.~Q.~Lu, Z.~G.~Huang, Class.Quantum Grav. \textbf{24}, 3799 (2007), arXiv:hep-th/0610188.
\bibitem{vikman1}V.~Mukhanov and A.~Vikman, JCAP \textbf{0602}, 004 (2006) arXiv:astro-ph/0512066v2.
\bibitem{Bonvin1}C.~Bonvin, C.~Caprini, R.~Durrer, Phys.\ Rev.\ Lett. \textbf{97}, 081303 (2006) 	 arXiv:astro-ph/0606584v2
\bibitem{Kang}J.~U.~Kang, V.~Vanchurin and  S.~Winitzki, Phys.\ Rev.\ D \textbf{76}, 083511 (2007) arXiv:0706.3994v2 [gr-qc].
\bibitem{vikman2}E.~Babichev, V.~Mukhanov and A.~Vikman, JHEP \textbf{02}, 101 (2008) arXiv:0708.0561v1 [hep-th].
\bibitem{Ellis}
 G.~Ellis, R.~Maartens, M.~MacCallum, Gen.\ Rel.\ Grav.\ \textbf{39}, 1651 (2007), arXiv:gr-qc/0703121.
\bibitem{Bonvin2}C.~Bonvin, C.~Caprini and R.~Durrer, arXiv:0706.1538v2 [astro-ph]

\bibitem{WMAP5} E. Komatsu et. al  arXiv:0803.0547v1 [astro-ph].
\bibitem{spergel_2006}D. N. Spergel et al. Astrophys. J. Suppl. 170, 377 (2007)
[astro-ph/0603449].
\bibitem{spergel_2003}D. N. Spergel et al. Astrophys. J. Suppl. 148, 175 (2003)
[astro-ph/0302209].
\bibitem{GTF1} L. P. Chimento, Phys.\ Rev.\ D \textbf{69},  123517 (2004).  arXiv:astro-ph/0311613v2
\bibitem{GTF2}
R. Yang, S. N. Zhang, Y. Liu JCAP \textbf{01}, 017 (2008).	arXiv:0802.2358v1 [astro-ph]
\bibitem{Hu_1998}
W.~Hu, Astrophys.J. \textbf{506}, 485 (1998)	arXiv:astro-ph/9801234v2.
\bibitem{Stojkovic}
D. Stojkovic, G. D. Starkman and R. Matsuo,  Phys.\ Rev.\ D \textbf{77}, 063006 (2008) [arXiv:hep-ph/0703246v2].
\end{thebibliography}
\end{document}